\documentclass[]{pasj01}
\Received{\today}
\Accepted{$\langle$acception date$\rangle$}
\Published{$\langle$publication date$\rangle$}
\SetRunningHead{Kurahara et al.}{Diffuse radio source in Abell 1060}

\usepackage{url}
\usepackage[switch,mathlines]{lineno}
\usepackage{natbib}
\bibpunct{(}{)}{;}{a}{}{,} 

\begin{document}

\title{Discovery of Diffuse Radio Source in Abell 1060}
\author{
Kohei \textsc{Kurahara}\altaffilmark{1},
Takuya \textsc{Akahori}\altaffilmark{1},
Aika \textsc{Oki}\altaffilmark{1,2},
Yuki \textsc{Omiya}\altaffilmark{3}, and
Kazuhiro \textsc{Nakazawa}\altaffilmark{3,4}
}

\altaffiltext{}{
$^1$Mizusawa VLBI Observatory, National Astronomical Observatory Japan, 2-21-1 Osawa, Mitaka, Tokyo 181-8588, Japan \\
$^2$Department of Astronomy, Graduate School of Science, The University of Tokyo, 7-3-1 Hongo, Bunkyo-ku, Tokyo 113-0033, Japan\\
$^3$Department of Physics, Nagoya University, Furo-cho, Chikusa-ku, Nagoya, Aichi 464-8601, Japan\\
$^4$Kobayashi-Maskawa Institute for the Origin of Particles and the Universe, Furo-cho, Chikusa-ku, Nagoya, Aichi 464-8601, Japan
}

\email{kohei.kurahara@nao.ac.jp}

\KeyWords{galaxies: clusters: individual (Abell 1060) --- galaxies: individual (NGC 3309, NGC 3311, NGC 3312)}

\maketitle

\begin{abstract}

Non-thermal components in the intra-cluster medium (ICM) such as turbulence, magnetic field, and cosmic rays imprint the past and current energetic activities of jets from active galactic neuclie (AGN) of member galaxies as well as disturbance caused by galaxy cluster mergers. Meter- and centimeter-radio observations of synchrotron radiation allow us to diagnose the nonthermal component. Here we report on our discovery of an unidentified diffuse radio source, named the Flying Fox, near the center of the Abell 1060 field. The Flying Fox has an elongated ring-like structure and a central bar shape, but there is no obvious host galaxy. The average spectral index of the Flying Fox is $-1.4$, which is steeper than those for radio sources seen at meter wavelength. We discussed the possibilities of radio lobes, phoenixes, radio halos and relics, and Odd Radio Circle (ORC). In conclusion, the Flying Fox is not clearly explained by known radio sources.
\end{abstract}


\clearpage

\section{Introduction}
\label{section1}

Galaxy clusters are formed by multiple mergers of groups and clusters of galaxies. While major mergers, where the mass of the two systems are comparable to each other, are dramatic and spectacular phenomena, minor mergers, in which a sub system falls by the gravity of the main system, are more frequent and play a role in disturbing the thermal and dynamical state of the intra-cluster medium (ICM) and facilitating chemical enrichment in galaxy clusters. Merger events impact on the galaxy evolution through galaxy mergers and ram-pressure stripping as head-tail galaxies as well as on the energy feedback of radio jets from AGN such as wide-angled-tailed (WAT) radio galaxies.

Hydrodynamic simulations of minor merger of galaxy clusters have suggested that the co-existence and mixing of multi-temperature ICM around the gravitational center of the main cluster \citep{2003ApJ...590..225C, 2004ApJ...612L...9F, 2016ApJ...821....6Z}. This phenomenon is called ``sloshing", where a relatively cool ICM is moving into the hotter ICM. The sloshing exhibits the temperature gap between the two ICM without raising shock front, because the sloshing motion is sub-sonic in the hot ICM. Actually, such a temperature gap has been observed in X-rays and the gap is called ``cold front" \citep[see,][for a review]{2007PhR...443....1M}.

The sloshing is thought to be one of the key mechanisms that link non-thermal components of the ICM such as turbulence, magnetic field, and cosmic rays. The sloshing motion results in forming turbulence, amplifying magnetic field, and driving turbulent re-acceleration (i.e. second-order Fermi acceleration) of cosmic rays. The turbulent acceleration models predict a steep ($\alpha \lesssim -1.5$ of $S\propto \nu^\alpha$) spectrum radio halos \citep[e.g.,][]{2001MNRAS.320..365B, 2001ApJ...557..560P, 2008Natur.455..944B, 2023A&A...672A..43C}, which have been found in some galaxy clusters \citep[e.g.,][]{2021A&A...650A..44B, 2021PASA...38...53D, 2021NatAs...5..268D, 2022MNRAS.515.1871R, 2023A&A...675A..51D}.

Centimeter and meter wavelength radio observations have provided fruitful information for understanding the non-thermal aspects of galaxies, AGNs, and galaxy clusters \citep[][for a review]{2018PASJ...70R...2A}. Cosmic-ray (re)-acceleration and magnetic-field amplification have been examined to explain classical radio halos, mini-halos, and relics, as well as faint radio fossil and phoenix which are thought to be aged radio sources but be brightening due to disturbance caused by subcluster mergers and/or injection of cosmic-rays from jets \citep[see][for a review]{2019SSRv..215...16V}. The fossils often exhibit an ultra-steep radio spectrum \citep[$\alpha \lesssim -2$; e.g.,][]{2017SciA....3E1634D, 2020A&A...634A...4M, 2020A&A...643A.172I}, and are more paid into attention, thanks to modern high sensitivity radio telescopes.

More recently, using the Australian Square Kilometre Array Pathfinder (ASKAP) \citep{2021PASA...38....9H} and the Giant Metrewave Radio Telescope (GMRT), \citet{2021PASA...38....3N} reported unidentified diffuse radio sources in the extragalactic space. They are called Odd Radio Circle (ORC). Only several ORCs have been reported and their spectral indice range from $-0.7$ to $-1.1$ \citep{2022MNRAS.513.1300N}. Several origins of ORCs have been discussed, including the shock wave from the central galaxy, the termination shock of galactic wind, the end-on-view double radio lobes, and the interactions between galaxies \citep{2021MNRAS.505L..11K, 2022MNRAS.513.1300N} as well as the throats of wormholes \citep[e.g.,][]{2021Univ....7..178K}. Cross-identification of host galaxies is essential to understand their origins.

In this letter, we report on our discovery of an unidentified diffuse radio source in the field of Abell 1060, using the GMRT. In Section 2, we briefly summarize the observation and our data reduction. The results and our discussions are described in Sections 3 and 4, respectively. Throughout the letter, we adopt the $\lambda$CDM cosmology giving $\sim 0.27$~kpc arcsec$^{-1}$ or $\sim 16$~kpc arcmin$^{-1}$ for the redshift to Abell 1060, $z=0.013$.

\section{Observation and Data Reduction}
\label{section2}

We mainly used the archival data taken with GMRT in December 2010. The project code is ddtB015. The data were recorded by the GMRT Hardware Backend (GHB) and the GMRT software correlator backend (GSB) in parallel. We analyzed the GSB data, where its center frequency and the bandwidth are 338~MHz and 33~MHz, respectively. The radio source 3C147 was used as the flux density and bandpass calibrator, and J1033-343 as the phase calibrator. The TGSS point-source catalog \citep{2017A&A...598A..78I} was used for self-calibration and peeling of compact sources. Moreover, the SPAM \citep[Source Peeling and Atmospheric Modeling;][]{2014ASInC..13..469I} pipeline was applied to achieve the direction-dependent calibration \citep[DDC;][]{2017A&A...598A..78I}. See \citet{2023PASJ...75S.138K} for details of the data reduction and the analysis.

To calculate the radio spectral index, we referred to GLEAM \citep{2017MNRAS.464.1146H}, TGSS \citep{2017A&A...598A..78I}, NVSS \citep{1998AJ....115.1693C} and RACS \citep{2020PASA...37...48M} data. Those data were taken from the NASA's SkyView facility (\url{http://skyview.gsfc.nasa.gov}) or the archival system. We also took WISE \citep{2010AJ....140.1868W} and DSS2-IR \citep{1990AJ.....99.2019L, 2000ASPC..216..145M} data from SkyView and XMM-Newton \citep{2001A&A...365L...1J} data from the archival system, for the cross-identification purpose. The X-ray data for a 64~ks exposure was analyzed following the previous method \citep{2023PASJ...75...37O}. The redshift was fixed at 0.013 and the metal abundance table of \citet{2009LanB...4B..712L} was used.

\section{Results}
\label{section3}

The top panel of figure~1 shows the 338~MHz total intensity (Stokes I) map near the center of Abell 1060. The brightest source possesses a bi-polar shape and its midpoint well-aligns with the center of the elliptical galaxy NGC 3309 (figure 2 left), suggesting that the source is radio jets ejected from AGN of NGC 3309. The peak brightness is 23 mJy beam$^{-1}$ or 245$\sigma$, where $\sigma$ = 0.094 mJy beam$^{-1}$ is the image rms noise level. The integrated flux density at 338 MHz is 334 mJy, which is in agreement with 288 mJy derived from the spectral index of $-1.12$ that is obtained from the power-law fit of the TGSS (874 mJy at 150 MHz), GLEAM (413 mJy at 200 MHz), and NVSS (61 mJy at 1400 MHz) data.

\begin{figure}[tp]
\begin{center}
\FigureFile(80mm,80mm){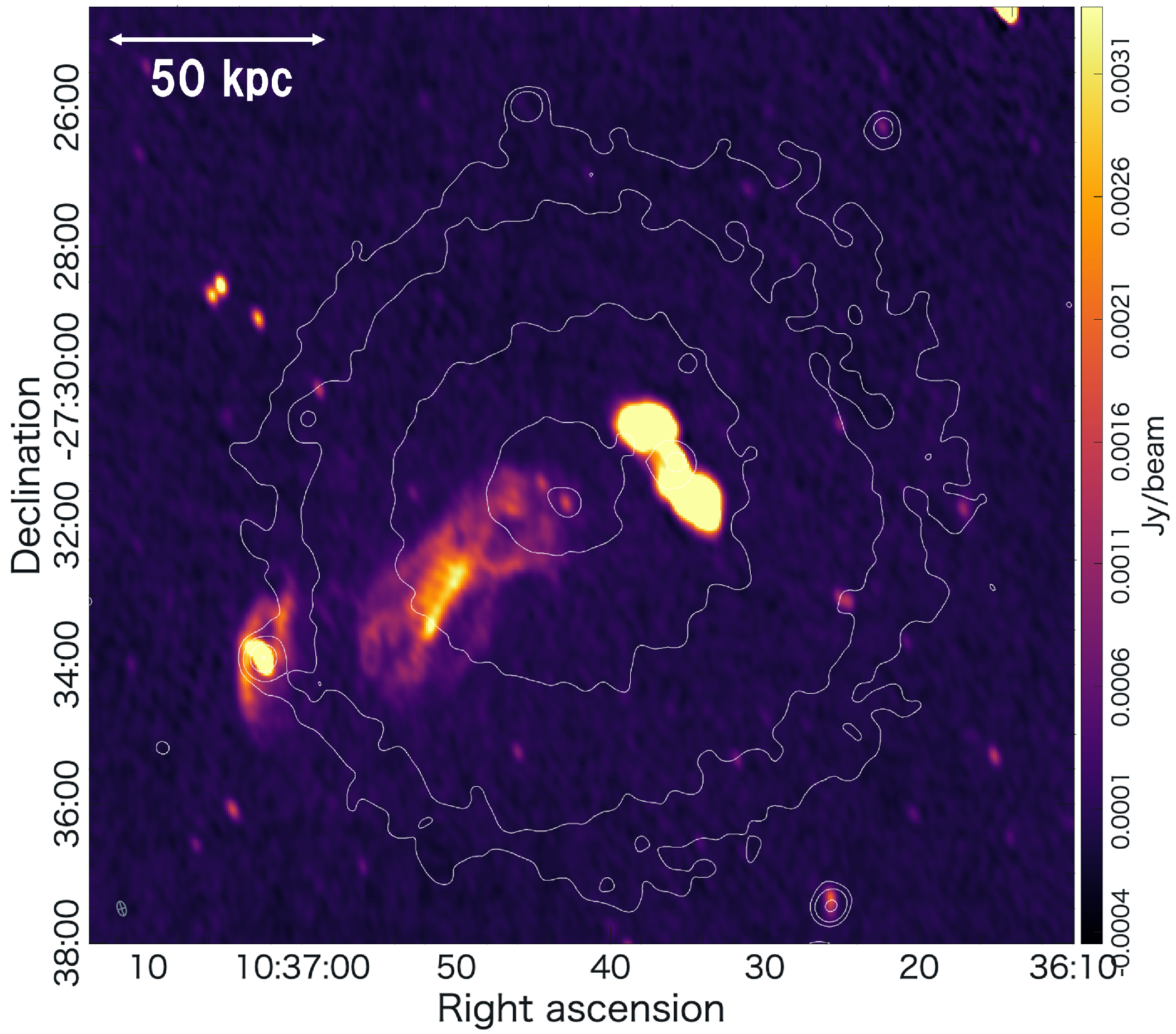}
\FigureFile(80mm,80mm){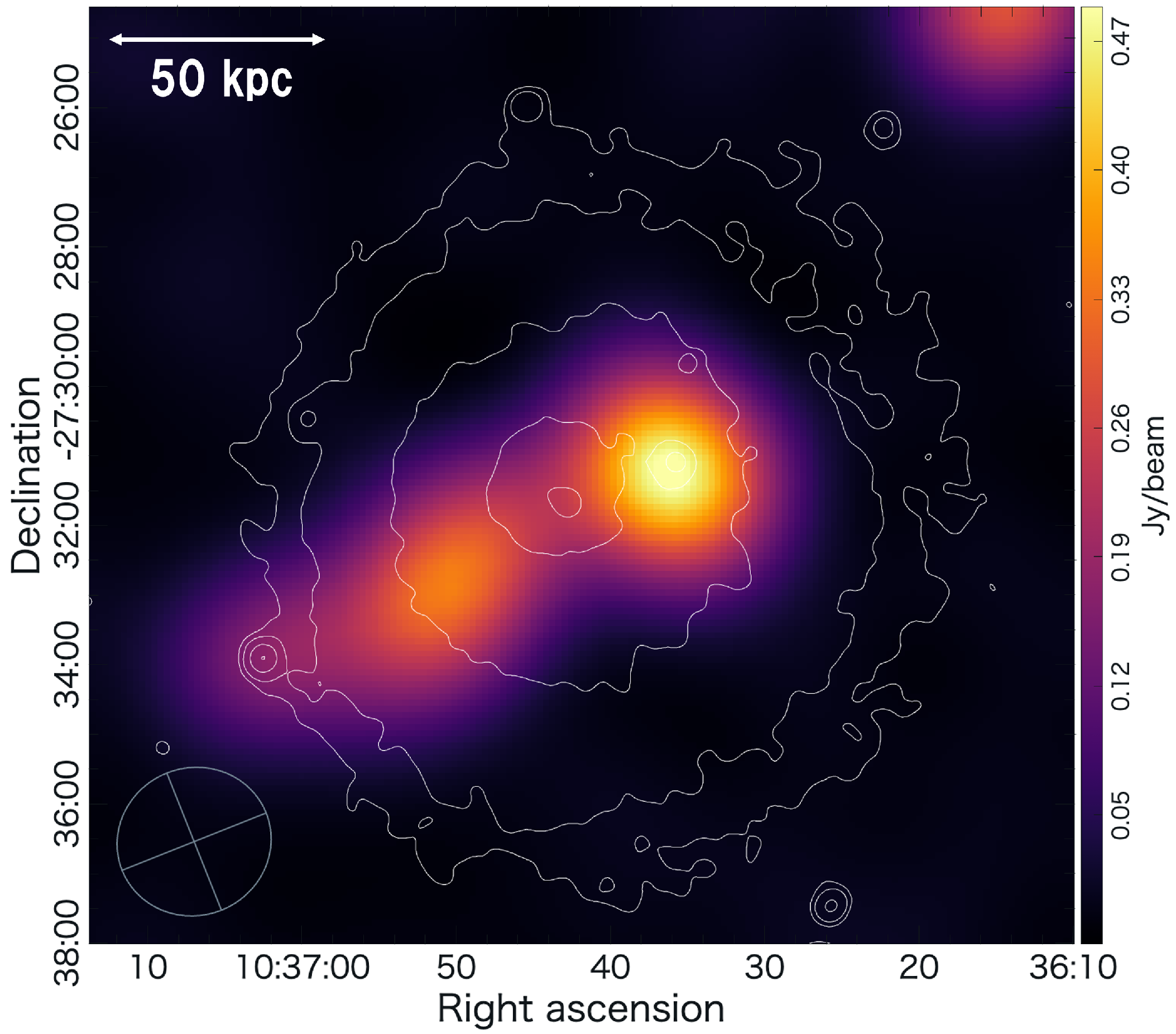}
\end{center}
\caption{The GMRT 338 MHz (top) and GLEAM 200 MHz (bottom) images of the Abell 1060 center. The white contours are the XMM-Newton X-ray surface brightness profile. Also drawn at the lower left are the synthesized beam sizes of $13''.1 \times 7''.2$ with the position angle of 17.5 degrees for uGMRT and $2'.2 \times 2'.1$ with $-$6.8 degrees.
}
\label{f01}
\end{figure}

\begin{figure*}[tp]
\begin{center}
\FigureFile(55mm,55mm){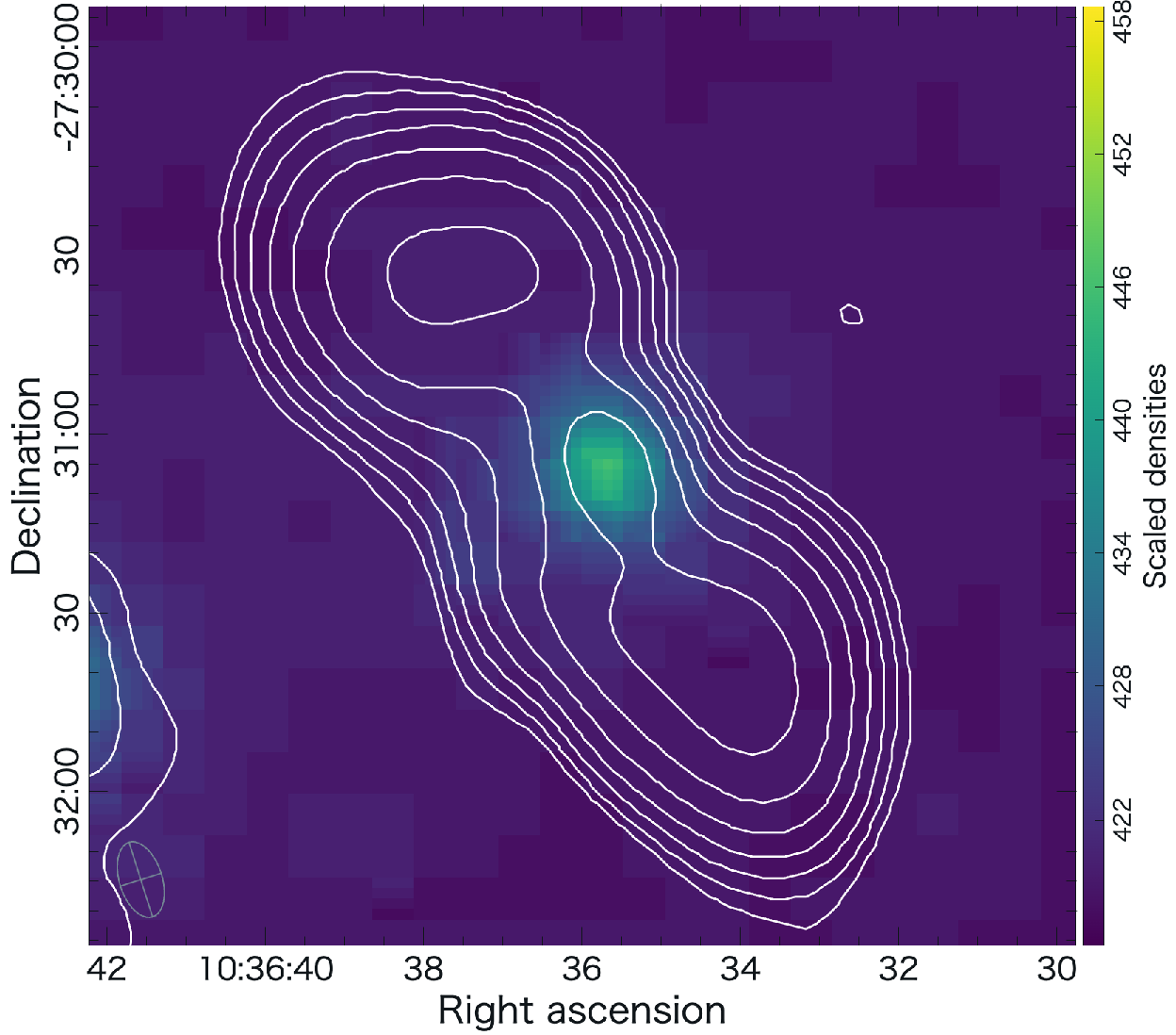}
\FigureFile(55mm,55mm){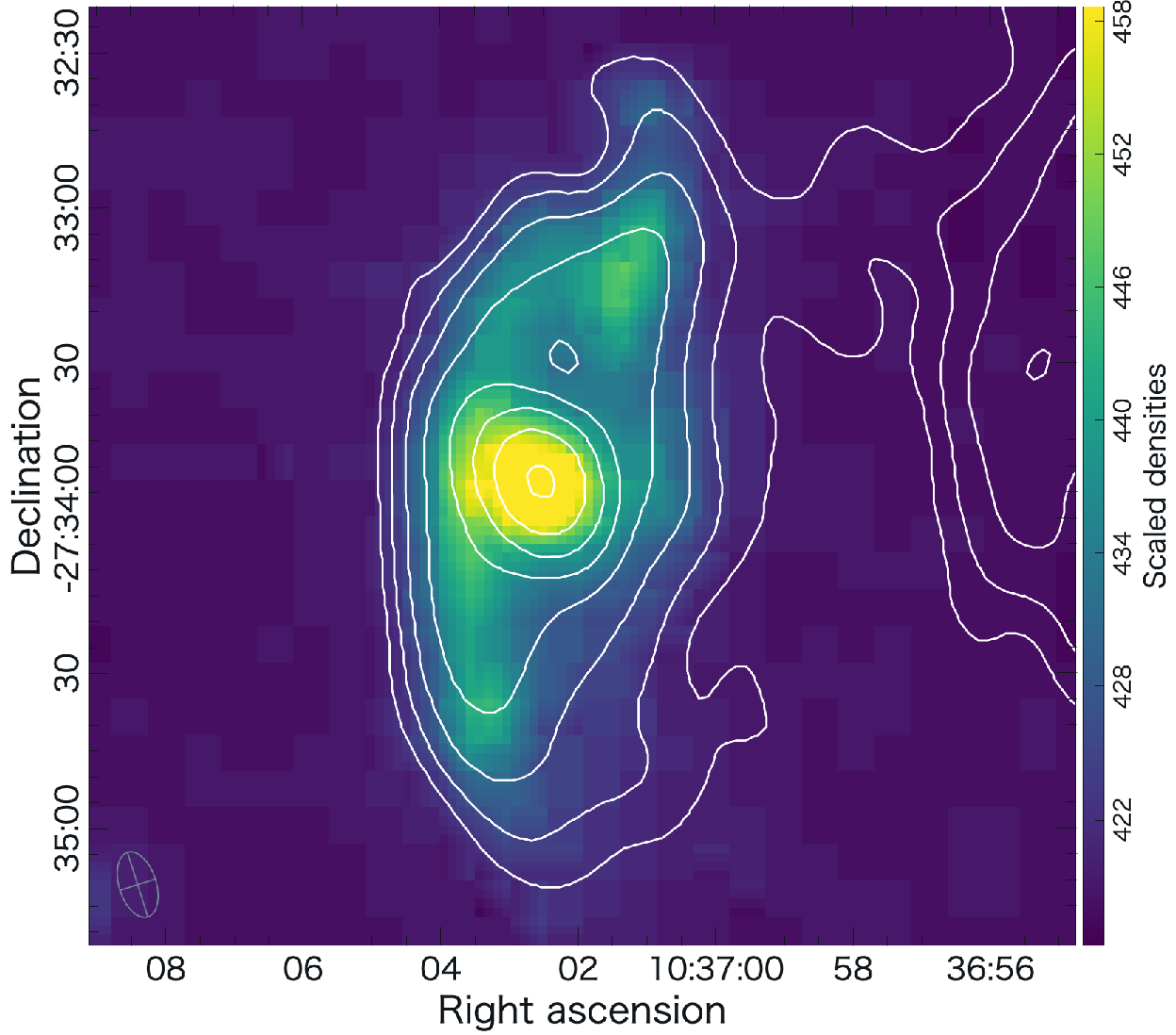}
\FigureFile(55mm,55mm){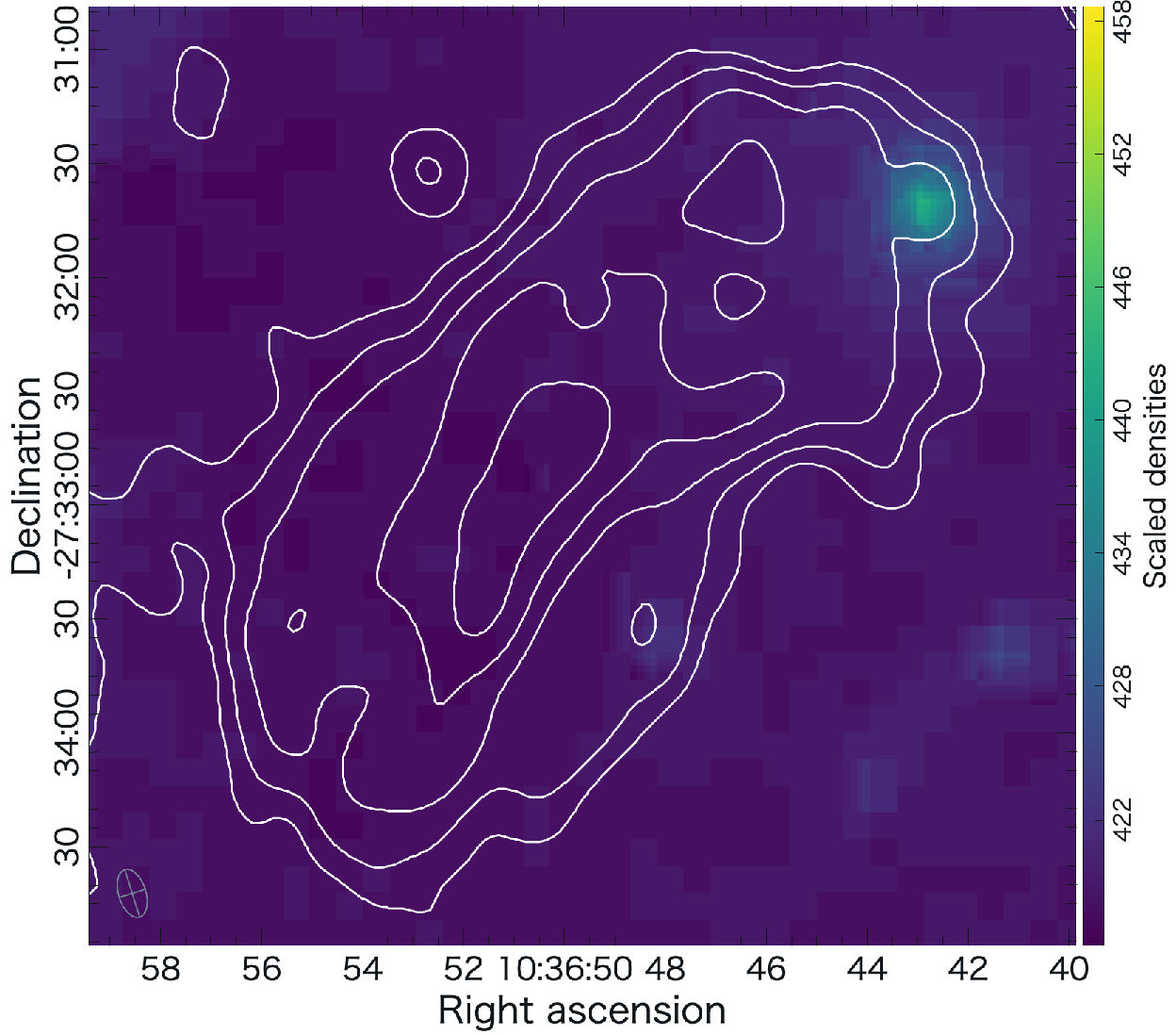}
\FigureFile(55mm,55mm){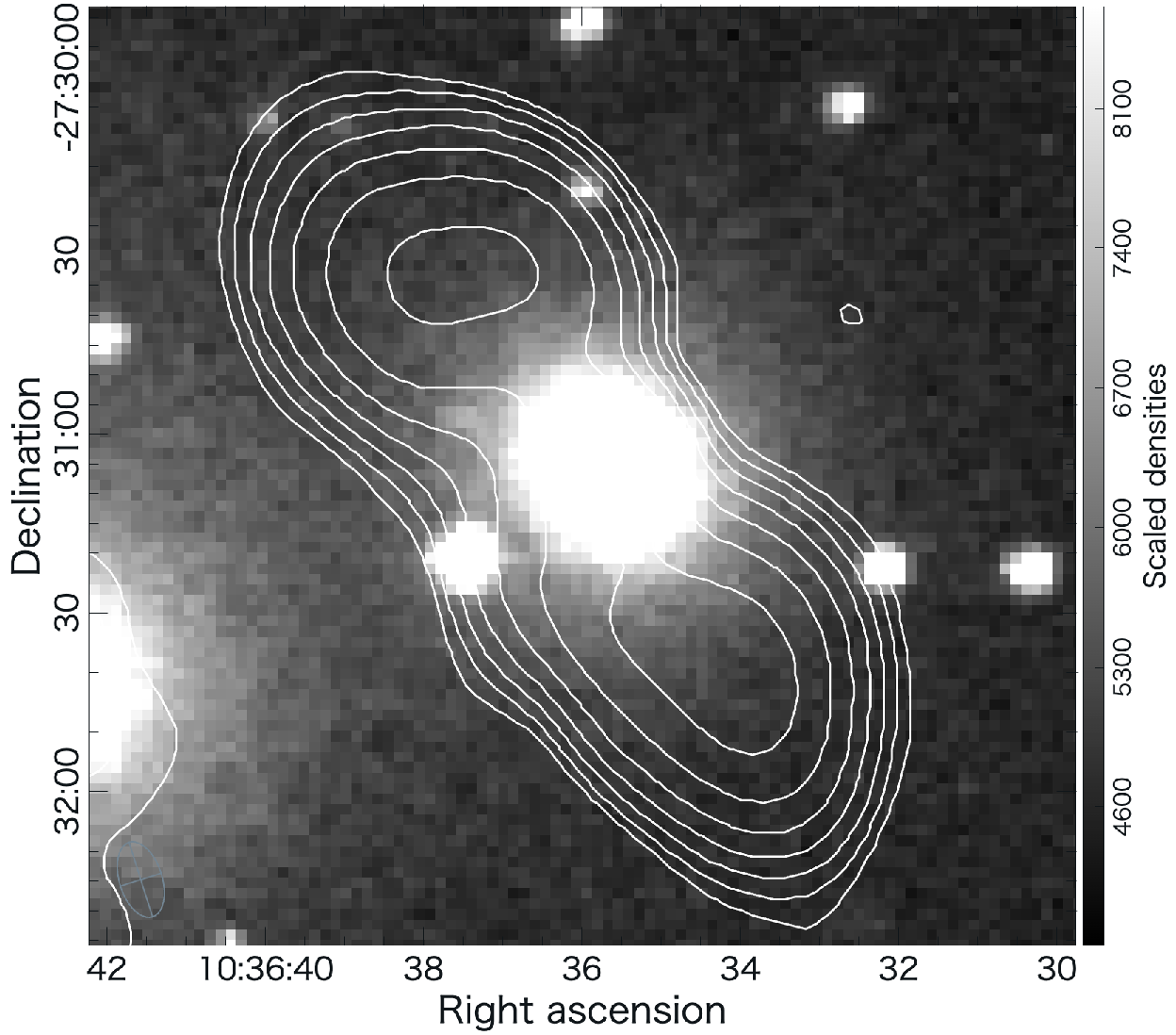}
\FigureFile(55mm,55mm){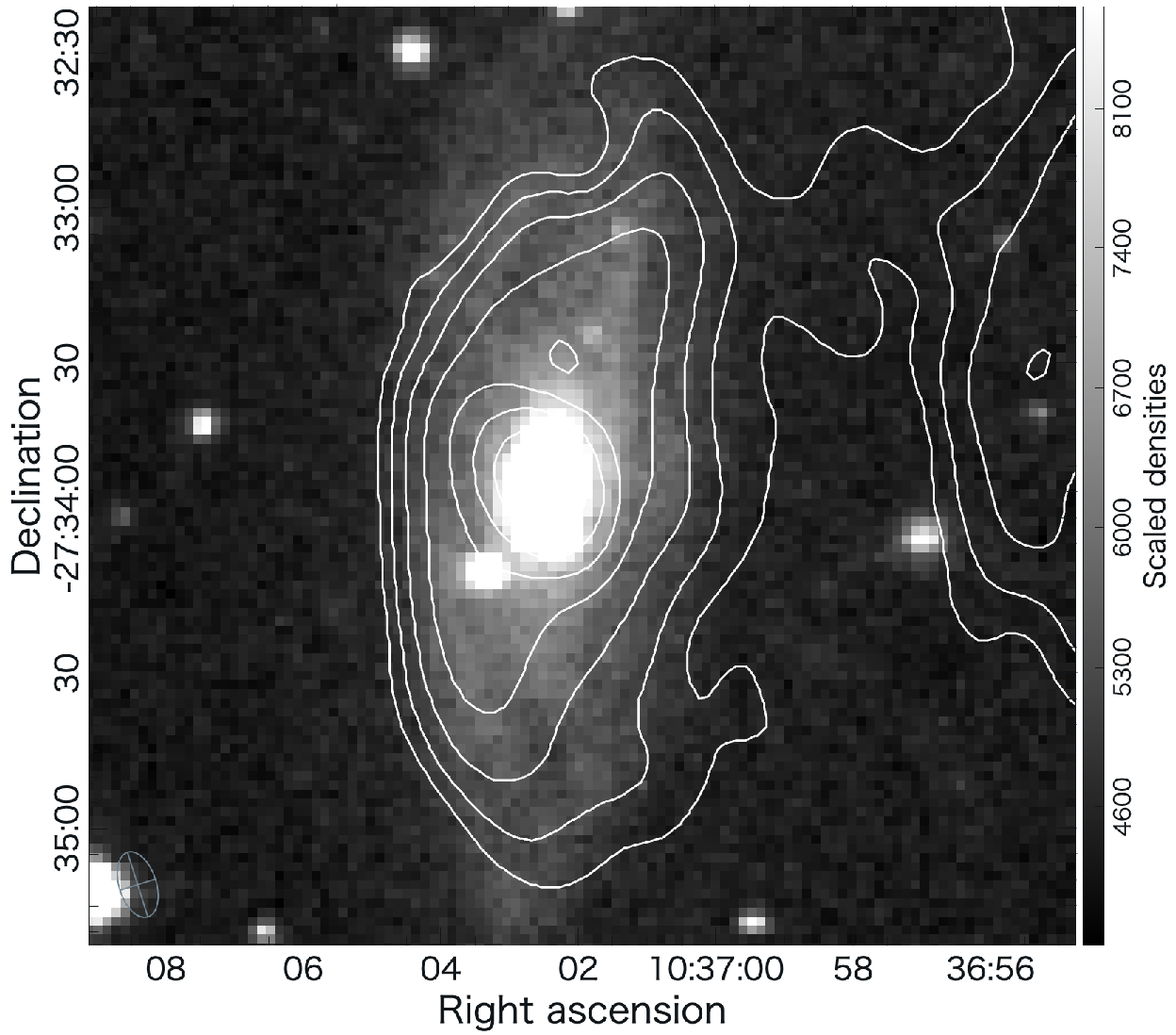}
\FigureFile(55mm,55mm){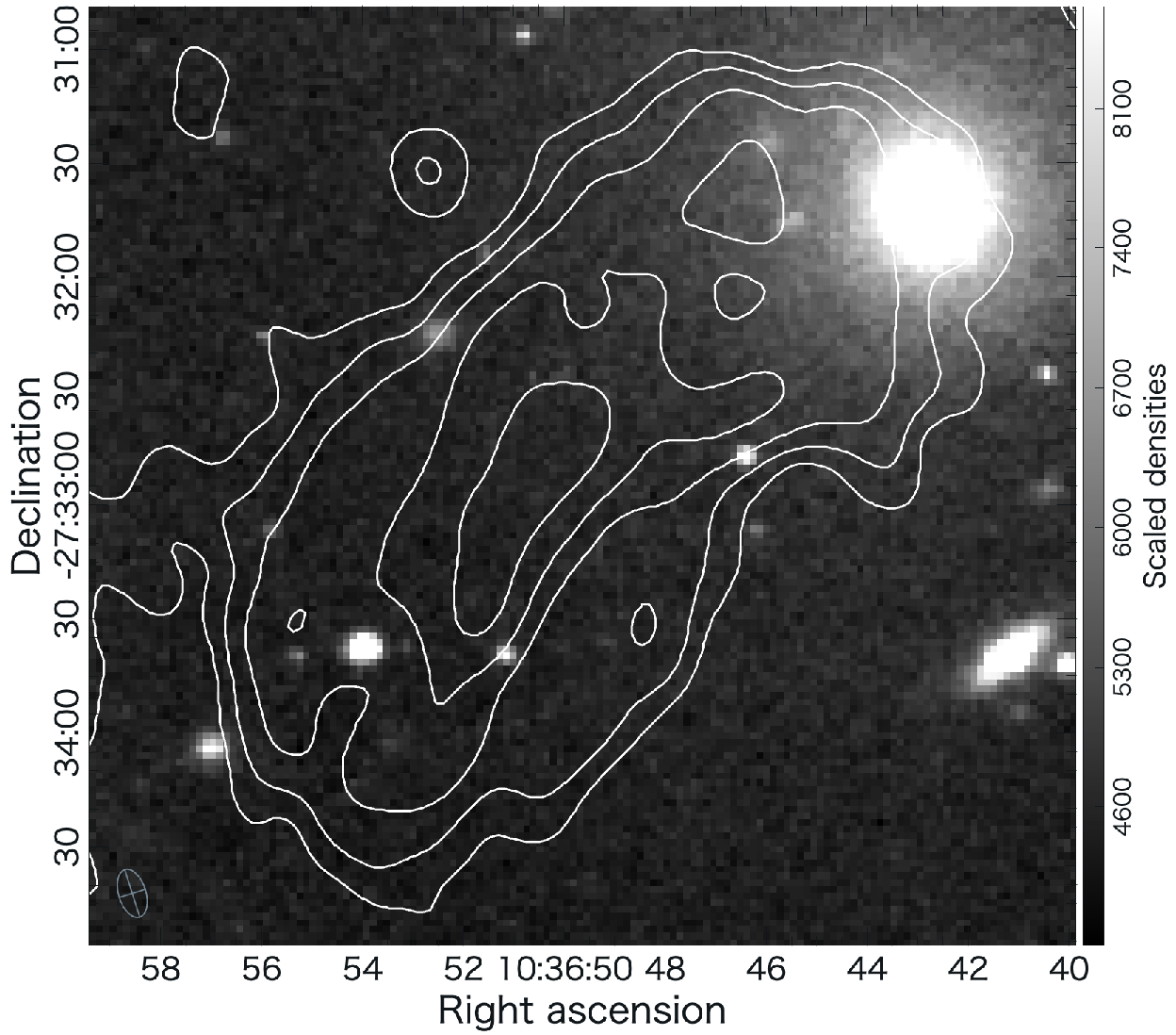}
\end{center}
\caption{The WISE 12 micron images (top) and DSS2-IR optical images (bottom) overlaid with the GMRT radio contours. The left, middle, and right panels show the zoom-in views around NGC 3309, NGC 3312, and NGC 3311, respectively.
}
\label{f02}
\end{figure*}

The second brightest radio source is located at the south-east part of the image. The peak brightness is 4.3  mJy beam$^{-1}$ or 45$\sigma$. The brightest center is associated with the center of the spiral galaxy NGC 3312 and an elongated ring shape matches with its arms (figure 2 middle), suggesting that the source is the diffuse synchrotron emission of NGC 3312. The integrated flux density at 338 MHz is 106 mJy, consistent with 94 mJy or the spectral index of $-0.48$ obtained from the TGSS (173 mJy at 150 MHz), GLEAM (95 mJy at 200 MHz), and NVSS (49 mJy at 1400 MHz) data.

With the above confirmation of the validity of our synthesized image and the flux scale, we find with GMRT that there is a spectacular diffuse radio source near the center of the image. This source shows an elongated ring-like shape with the middle part brightening strongly in a straight bar. We hereafter call it {\it the Flying Fox} based on its silhouette. The head of the Flying Fox is aiming at south-west. Its major axis (or the wingspan) is $\sim 235$~arcsec ($\sim 66$ kpc) and its minor axis (head-and-body length) is $\sim 100$~arcsec ($\sim 28$~kpc). The brightest bar (the inner right-wing to the chest) has a brightness of 3.2 mJy beam$^{-1}$ (34$\sigma$) and the outer ring has $\sim 0.65$~mJy beam$^{-1}$ (7$\sigma$), both confirming solid detection. The elliptical galaxy NGC 3311 overlaps at the north-western edge of the ring structure (or the left-wing tip), where a point-like excess is associated with the center of NGC 3311 (figure 2 right).

We confirmed that the Flying Fox was detected in MWA GLEAM as an unresolved source with the flux density of $431 \pm 23$ mJy at 200 MHz (figure~1 bottom). With our detection of the integrated flux density at 338 MHz, $208 \pm 21$ mJy, the average spectral index $\alpha$ at the meter-wavelength is $-1.4$, which is notably steep compared to the sources in the GLEAM catalog (typically $-0.77$). We confirmed that the extrapolated brightness of the radio peak is below the detection levels of RACS at 900 MHz and NVSS at 1400 MHz. There is a lack of the sensitivity with GMRT to split the data and obtain the in-band spectral index. Thus, the spectral index map was not obtained.

\section{Discussion}
\label{section4}

We discuss the possible origin of the Flying Fox. In conclusion, the Flying Fox is not clearly explained by known radio sources such as radio lobes and their fossils, diffuse cluster emissions, and ORCs. A future exploration of host galaxies with optical/IR telescopes will help understanding the origin of the Flying Fox. A future study of the spectrum-index map will provide the information of the age and evolution of the Flying Fox, while the polarization map will reveal the magnetic-field structure. Those information are also crucial to understanding the origin and nature of the Flying Fox.

\subsection{Remnant of Radio Lobes of Member Galaxies?}

The overall morphology of the ring-like structure and the central bar shape implies that the Flying Fox resembles radio lobes of AGN jets. However, we confirmed no clear host AGN and galaxy at the midpoint of the Flying Fox in radio/optical/IR/X-ray images. NGC 3311 connects to the ring-like structure and possesses radio-bright AGN core, so that one possibility is that the Flying Fox is a WAT-like radio jets from NGC 3311 and the bright bar is the hot spots. This scenario suggests that NGC 3311 is moving to the north-west direction to form a stripped tail or a remnant of radio lobes, where the cosmic-ray aging from the north-west to the south-east direction is expected and is needed to be consistent with the observed steep spectrum of the integrated flux density at meter wavelength. This scenario would be supported by X-ray (figure 3); the temperature and metal abundance maps implies a sloshing structure and the Flying Fox is aligned with the high-temperature and metal-rich region, that seems to be a tail-like structure from NGC 3311.

The steep spectrum implies that the Flying Fox is an aged source. One may consider that the Flying Fox is an old radio lobe made by the past jet activity of NGC 3311, and NGC 3311 was escaped from the radio lobe with a relative velocity to the ICM. Such a relative velocity can be seen in merger simulations including sloshing; dark matter and galaxies are collision-less components and are moving forward from the ICM. Assuming a relative velocity of 100~km s$^{-1}$ on the sky-plane, the distance between NGC 3311 and the center of the ring, 24~kpc, gives the travel time of 240 Myr. This time scale is broadly consistent with radio fossils discussed in the literature \citep[e.g.,][]{2019SSRv..215...16V}, although it is difficult to precisely estimate such a relative velocity between the ICM and the galaxy.

\begin{figure}[tp]
\begin{center}
\FigureFile(80mm,80mm){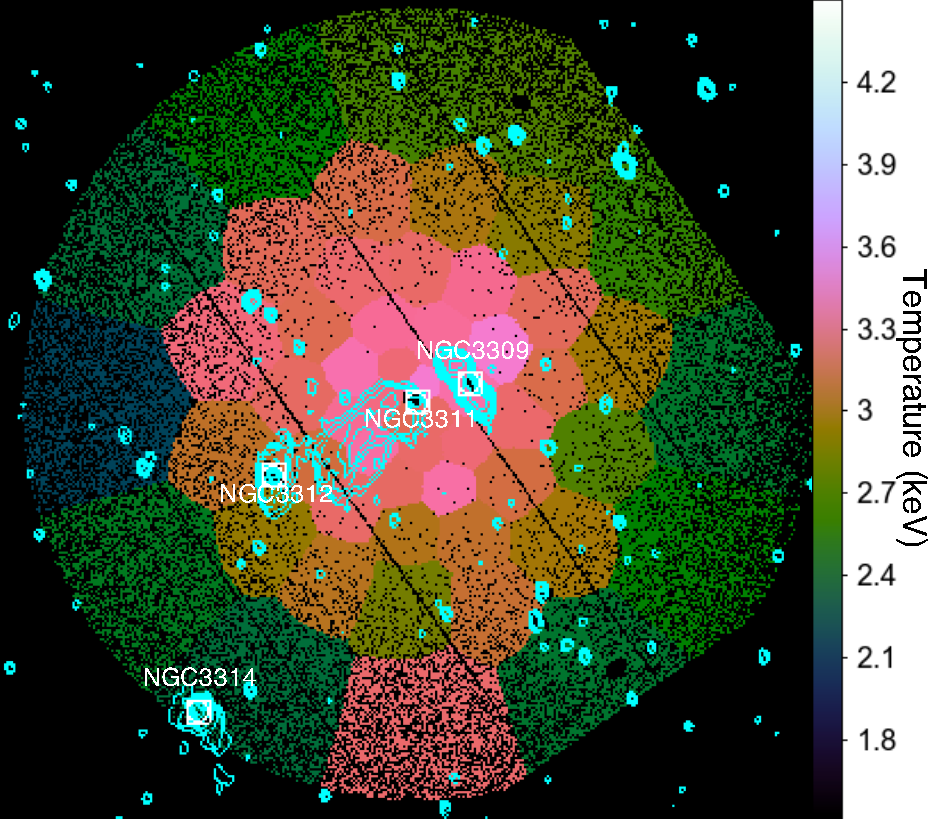}
\FigureFile(80mm,80mm){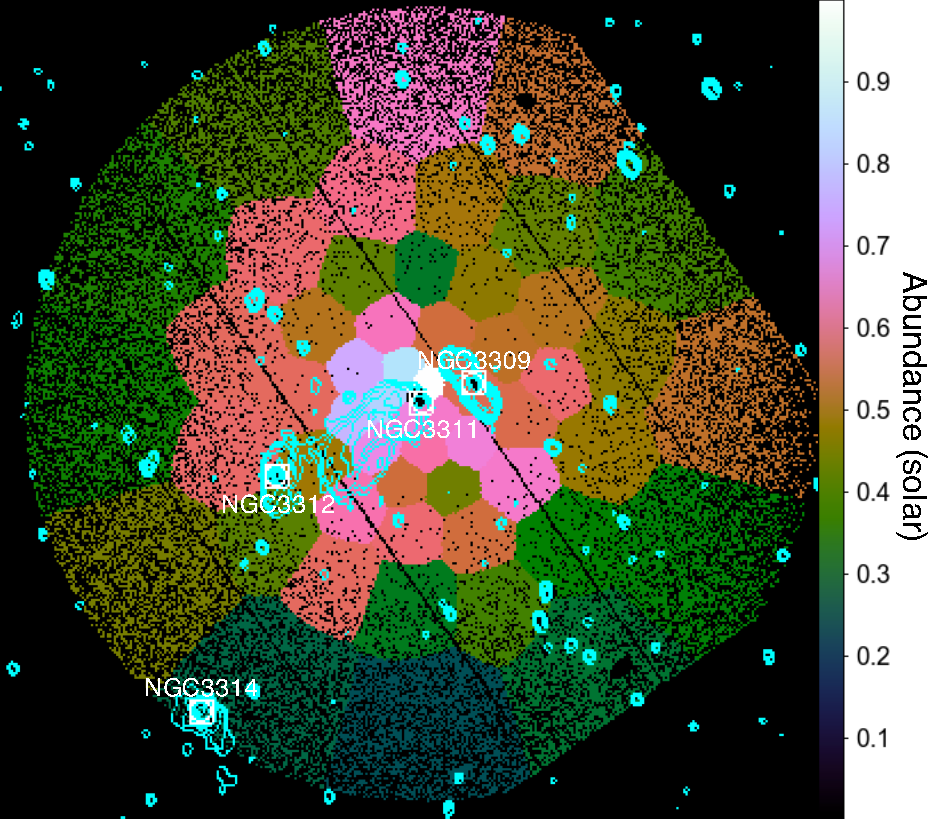}
\end{center}
\caption{The XMM-Newton temperature (top) and metal abundance (bottom) maps overlaid with the GMRT radio contours.}
\label{f03}
\end{figure}

The motion of NGC 3311 toward the north-west direction is, however, a controversy. NGC 3311 is located around the center of the line-of-sight velocity distribution of member galaxies \citep[see][]{2022A&A...668A.184H} and around the center of the X-ray surface brightness (figure \ref{f01}). Interestingly, the MeerKAT HI observation revealed that NGC 3312 and NGC 3314, which is located at 222 kpc south of NGC 3312 (see figure \ref{f03}), have HI tails to the south-west direction, suggesting that they are stripped gas from the galaxies both moving to the north-east direction. This motion of NGC 3312 does not align with the hypothetical motion of NGC 3311 toward the north-west direction. Note that it is difficult to examine the motion of NGC 3311 from HI observation, because NGC 3311 is a HI-poor elliptical galaxy and no HI emission has been detected \citep[][]{2022A&A...668A.184H}.

Leaving from NGC 3311, another potential origin of the Flying Fox is NGC 3309. If a relative motion between the ICM and NGC 3309, likely caused by a sloshing, is a clockwise or counter-clockwise circular motion around the gravitational center close to NGC 3311, NGC 3309 is expected to be located near the current position of the Flying Fox in the past. Thus, the Flying Fox may be a remnant of the radio lobe made by past jets from NGC 3309. However, the current jets and radio lobes associated with NGC 3309 do not indicate clear stripped tails, suggesting that NGC 3311 does not have such a circular motion. Note that Abell 1060 is classified into the Bautz-Morgan type III. Neither NGC 3311 nor NGC 3309 is the cluster galaxy which dominates the cluster center.

Finally, let us also discuss the scenario that the Flying Fox was formed by NGC 3312. \citet{2022A&A...668A.184H} revealed that NGC 3312 is located at the nearside in Abell 1060 and it is moving toward us according to its blue-shift with respect to the redshift distribution of Abell 1060. They suggest that NGC 3312 already experienced the passage of the pericenter of the Abell 1060 gravitational potential, so that the HI tails of NGC 3312 were formed when NGC 3312 passed through the dense ICM on its orbit. The Flying Fox may be a remnant of this disturbance at the passage. A weakness of this scenario is no clear enhancement of X-ray brightness and temperature. It is also possible that NGC 3312 is before the pericenter passage and is located at the backside in Abell 1060. In this case, the Flying Fox is not related to NGC~3312.

\subsection{Radio Relic or Halo?}

No detection of host galaxy suggests that the Flying Fox is classical cluster diffuse emission such as radio relic and halo. Radio mini-halo is also classified into the cluster diffuse radio source, but a mini halo  is often associated with the central brightest cluster galaxy (BCG). There is no galaxy around the center of the Flying Fox and no dominant galaxy in Abell 1060. Those natures disfavor that the Flying Fox is a radio mini-halo.

There is a well-known group of galaxies, HGC 48, located at about 500~kpc away to the north-east from NGC 3311. The redshift of HGC 48 is $z=0.094$, close to the redshift of Abell 1060. If HGC 48 is a post-merging group and it disturbed the ICM of Abell 1060 in the past, the Flying Fox may be a remnant of this event. There are both scenarios of radio relic and radio halo described below. Note that there is no data of the polarization fraction, which is often used to distinguish between these two.

Based on the X-ray observations as shown in figure 3, there is no notable edge of the temperature distribution associated with the Flying Fox. Therefore, it is difficult to explain the Flying Fox as an edge-on view of a radio relic. The Flying Fox is located at only $\sim 24$~kpc away from the cluster center. This location is very different from the nature that radio relics are often seen in the outskirts of clusters. Meanwhile, if HGC 48 is orbiting in the Abell 1060 gravitational potential, and it formed the Flying Fox when the projected position was near the Abell 1060 center on the sky plane, the Flying Fox should be located at the nearside or farside in Abell 1060. This orbital motion does not significantly disturb the central part of Abell 1060, showing no clear temperature jump in the projected X-ray map.

Another scenario which links to radio halo is that HGC 48 collided through the central part of Abell 1060. The collision could form the turbulence that can stretch and amplify magnetic fields and accelerate cosmic-rays by turbulent (re)-acceleration, resulting in radio halo. Although the spectral index of the Flying Fox is steeper than those of classical halos, recent observations have reported the steep-spectrum ones (see Introduction). However, the elongated morphology of the Flying Fox is different from those of radio halos. The ring-like structure is not expected in general for radio halos made by turbulent (re)-acceleration.

\subsection{Odd Radio Circle?}

It is interesting to argue whether the Flying Fox belongs to a category of ORC. A clear difference between the Flying Fox and ORCs is that the Flying Fox is very elongated while ORCs exhibit circular morphology. Another discrepancy, but which is not yet statistically significant, is that the Flying Fox is a steep spectrum source while ORCs are not. 

One possibility is that the Flying Fox is the aged ORC that was stretched toward the east-west direction by the gas sloshing with NGC 3311 toward the north-west direction, or it was compressed along the north-south direction by the ram pressure with aligning the motions of NGC 3312 and NGC 3314. Whichever the case is, no host galaxy is present and thus the origin is unclear.

\begin{ack}
This work was supported in part by JSPS KAKENHI Grant Numbers, JP21H01135(TA). AIPS is produced and maintained by the National Radio Astronomy Observatory, a facility of the National Science Foundation operated under cooperative agreement by Associated Universities, Inc.
\end{ack}

\bibliographystyle{aa}
\bibliography{Abell1060_FlyingFox}

\end{document}